\documentclass[a4paper,11pt]{article}
\pdfoutput=1 

\usepackage{jcappub} 

\usepackage[T1]{fontenc} 
\usepackage[utf8]{inputenc}
\usepackage{amsfonts}
\usepackage{lmodern}
\usepackage{amsthm,graphicx}
\usepackage{epsfig}
\usepackage[normalem]{ulem}
\usepackage{latexsym, amssymb} 
\usepackage{amsmath}

\DeclareMathAlphabet\mathbfcal{OMS}{cmsy}{b}{n}

 \usepackage{xcolor}
 \newcounter{attnctr} 

\title{\boldmath A non-parametric consistency test of the $\Lambda$CDM model with Planck CMB data}

\author[a]{Amir Aghamousa,}
\author[b,c]{Jan Hamann,}
\author[a,d]{Arman Shafieloo}

\affiliation[a]{Korea Astronomy and Space Science Institute, Daejeon, 305-348 Korea}
\affiliation[b]{School of Physics, The University of New South Wales, Sydney NSW 2052, Australia}
\affiliation[c]{Sydney Institute for Astronomy, School of Physics, The University of Sydney, Sydney NSW 2006, Australia}
\affiliation[d]{University of Science and Technology, Daejeon 34113, Korea}

\emailAdd{amir@aghamousa.com}
\emailAdd{jan.hamann@unsw.edu.au}
\emailAdd{shafieloo@kasi.re.kr}

\abstract{

Non-parametric reconstruction methods, such as Gaussian process (GP) regression, provide a model-independent way of estimating an underlying function and its uncertainty from noisy data. We demonstrate how GP-reconstruction can be used as a consistency test between a given data set and a specific model by looking for structures in the residuals of the data with respect to the model's best-fit. Applying this formalism to the \textit{Planck} temperature and polarisation power spectrum measurements, we test their global consistency with the predictions of the base $\Lambda$CDM model.  Our results do not show any serious inconsistencies, lending further support to the interpretation of the base $\Lambda$CDM model as cosmology's gold standard.

}

\begin{document}
\maketitle
\flushbottom

\section{Introduction}~\label{sec:intro}
The \textit{Planck} mission's~\cite{Adam:2015rua} measurement of the cosmic microwave background's anisotropies is currently probably the most powerful source of cosmological information. 
One of the main results of \textit{Planck} is a confirmation of the base $\Lambda$CDM model as the simplest phenomenologically viable cosmological model.  This conclusion is based on (i) extensive testing for {\it known} systematic effects at all stages of Planck data analysis from time-ordered data to parameter estimation~\cite{Aghanim:2015xee}, and (ii) a wide-ranging exploration of the space of physically motivated models~\cite{Ade:2015xua,Ade:2015lrj,Ade:2015rim}.

Nonetheless, while the overall goodness-of-fit of the \textit{Planck} data to base $\Lambda$CDM is reasonable, one cannot entirely rule out the possibility of (i) remaining {\it unknown} systematics, or (ii) the existence of a model that would be preferred over base $\Lambda$CDM by some measure such as the Bayesian model evidence.

Here we argue that parametric methods are not amenable to finding such effects and suggest a test of the consistency of \textit{Planck} data with base $\Lambda$CDM based on a non-parametric method which can be used to detect possible hidden patterns in the residuals of \textit{Planck}'s angular power spectra.  A statistically significant discovery of such a pattern would indicate problems with the model or our understanding of the data.

This test should be considered another item in the long list of challenges to base $\Lambda$CDM.  Naturally, owing to any method's inherent limitations, passing such a test cannot be an ultimate proof of base $\Lambda$CDM being correct, but would still bolster confidence in our present understanding of the physics governing the history of the Universe being adequate.

In particular, we employ Gaussian Process regression (see, e.g., \cite{Rasmussen:2006gp, Heitmann2010a, Heitmann2010b,Heitmann2011,Arman2012gp,Seikel2012gp}) as a non-parametric regression tool to quantify the consistency of \textit{Planck} 2015 temperature and polarisation angular power spectrum measurements with base $\Lambda$CDM.  

Previous cosmology model-independent and non-parametric analyses of the \textit{Planck} 2013 data in Refs.~\cite{Aghamousa2015b,Aghamousa2015c} identified an inconsistency between the concordance model and the data at a $2$-$3\sigma$ level using different statistical approaches. However, further analysis showed that a systematic effect in the 217~GHz spectrum was mainly responsible for the inconsistency~\cite{Hazra:2014hma}.

This paper is organized as follows: in Section~\ref{sec:method_data} we explain the characteristics of the angular power spectrum data sets used in this work, elaborate the methodology of our Gaussian process-based consistency test and demonstrate its ability to find certain classes of patterns.  In Section~\ref{sec:results_discussion} we show and discuss the results of applying the test to \textit{Planck} angular power spectrum data, and finally in Section~\ref{sec:conclusion} we present our conclusions.

\section{Data and Methodology}~\label{sec:method_data}
In this section we introduce the CMB angular power spectrum data sets used in this analysis. We also briefly review the theory behind Gaussian process regression (see~\cite{Rasmussen:2006gp}) and explain the way we utilize it to test the consistency of the $\Lambda$CDM model with CMB angular power spectrum data.
\subsection{Data}~\label{sec:data}
In this analysis we consider measurements of the CMB's TT, TE and EE angular power spectra released by the {\it Planck} collaboration in 2015.  We extract the data vectors and corresponding covariance matrices from the \texttt{plik\_lite} likelihood available on the \textit{Planck} Legacy Archive.\footnote{\url{http://pla.esac.esa.int}}  These data combine the information from {\it Planck}'s frequency channels and are already marginalised over foreground contributions.

\begin{figure}[ht]
\begin{center}
 \includegraphics[height=.57\columnwidth,angle=270]{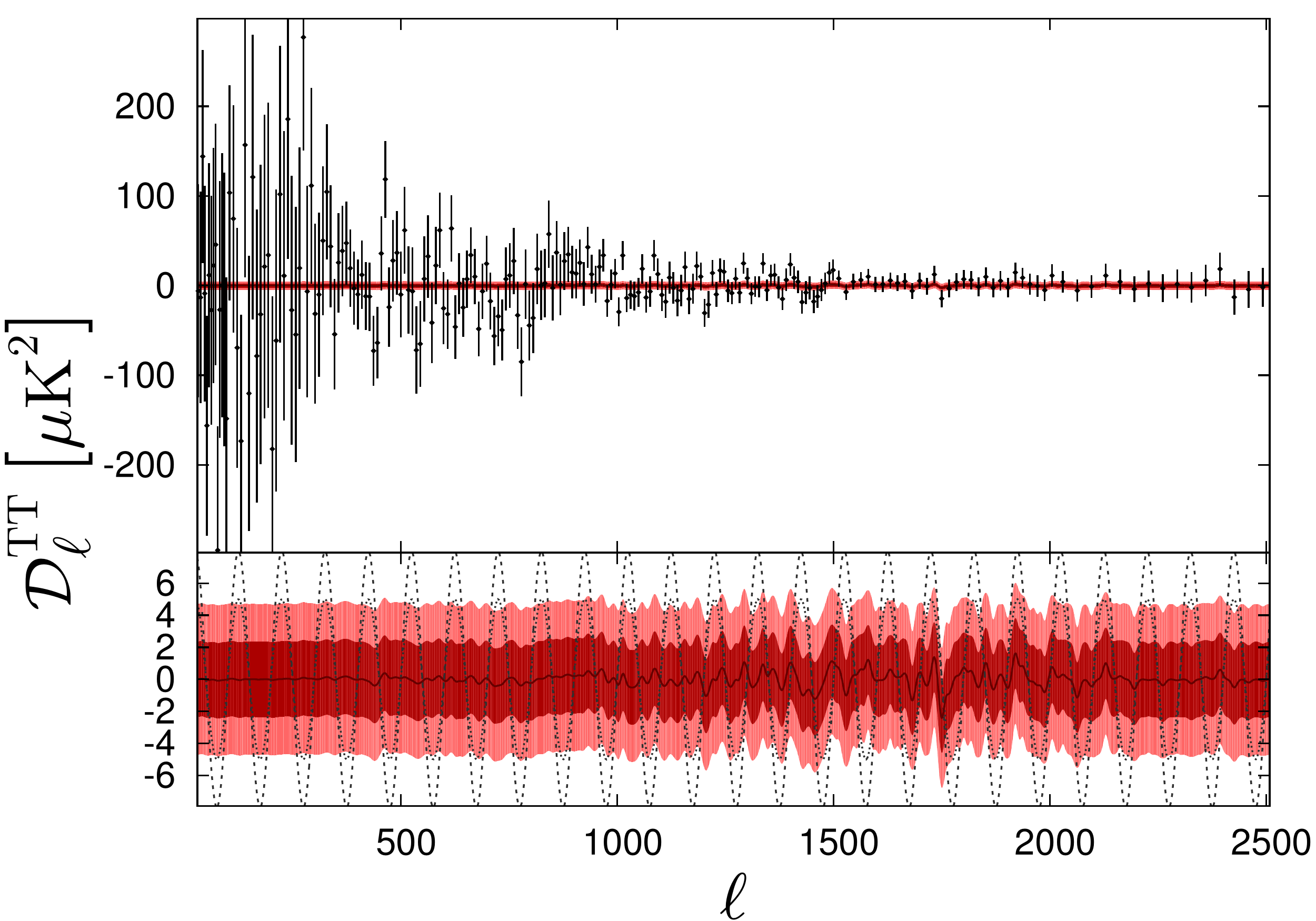}
 \includegraphics[height=.57\columnwidth,angle=270]{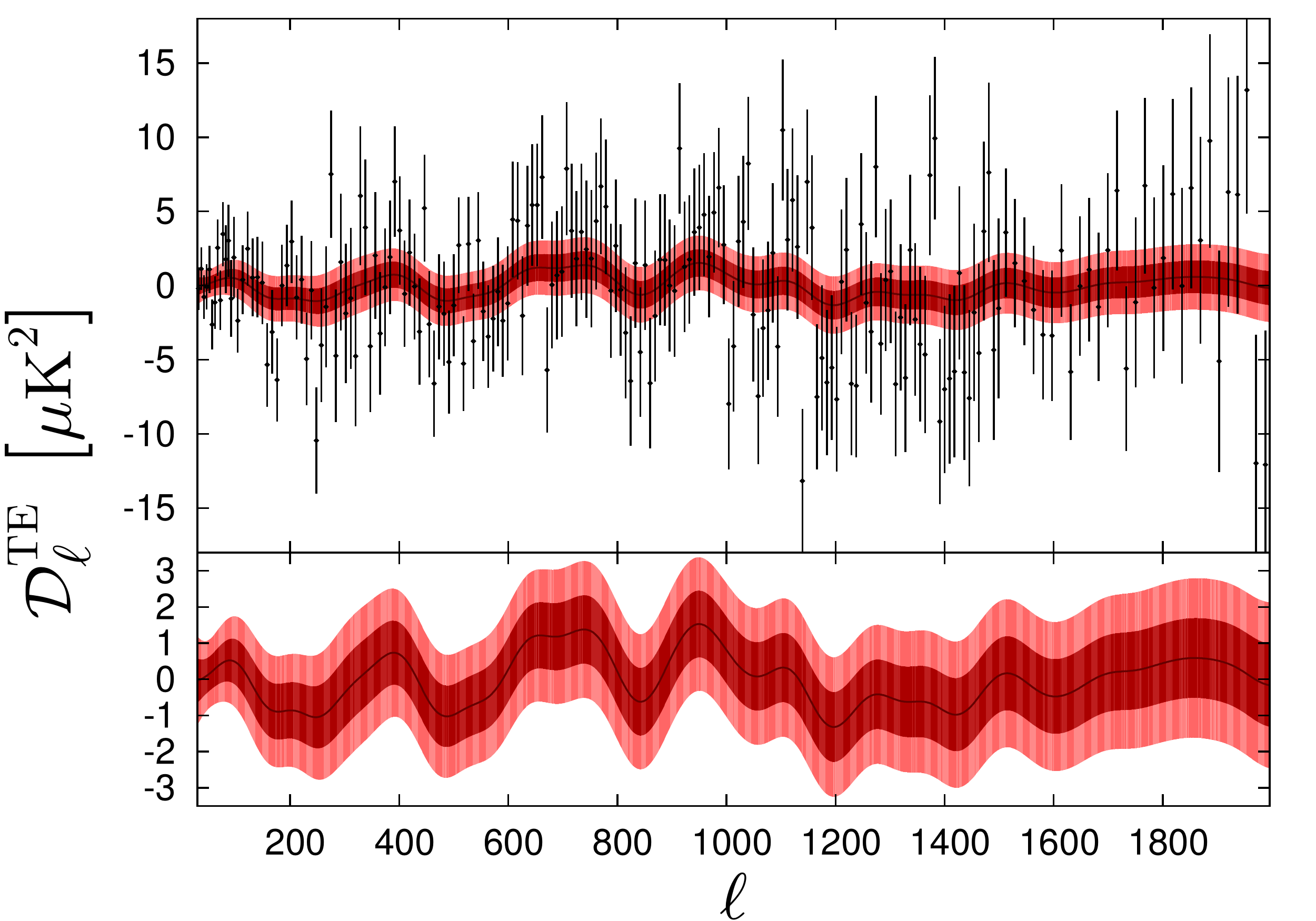}
 \includegraphics[height=.57\columnwidth,angle=270]{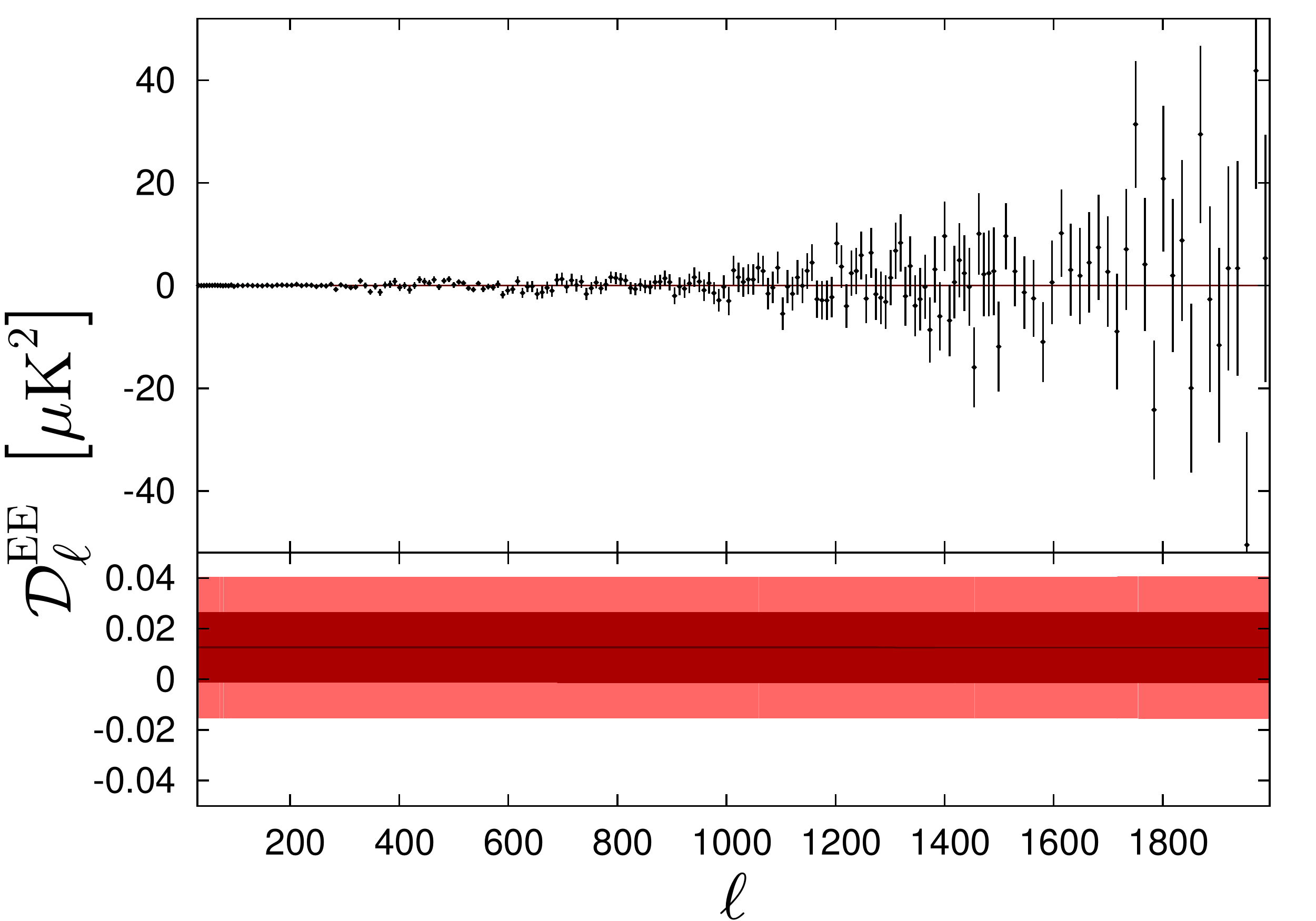}
 \caption{\label{fig:residuals_GP} These figures show the residuals $\mathcal{D}_\ell$ of the base $\Lambda$CDM model best fit with respect to the \textit{Planck} TT (\textit{top}), TE (\textit{centre}) and EE (\textit{bottom}) angular power spectrum data and the means of the respective maximum-likelihood Gaussian process regression functions (\textit{dark red lines}) along with their one and two standard deviation uncertainty bands. For better clarity, we provide a zoomed-in version of the Gaussian process function in the lower panels of the three plots.  The dotted and dashed grey lines indicate the magnitude of the fake signal for the examples considered in Section~\ref{sec:GPconsistency}.}
 \end{center}
\end{figure}

The TT power spectrum data cover the multipole range $30 \leqslant \ell \leqslant 2508$ and are given in $215$ bins with $\ell$-dependent widths, increasing from $\Delta \ell = 5$ at $\ell_\mathrm{min}$ to $\Delta \ell = 33$ at $\ell_\mathrm{max}$.  The EE and TE spectra are provided in $199$ bins of widths $5 \leq \Delta \ell \leq 17$ over a range of $30 \leqslant \ell \leqslant 1996$.  We plot the residuals of the data with respect to the $\Lambda$CDM model's best-fit spectra in Figure~\ref{fig:residuals_GP}.
Even if the data were unbinned, in this multipole range the data follow to an excellent approximation a Gaussian likelihood distribution~\cite{Aghanim:2015xee}.  This motivates our use of Gaussian process regression, as described in the following subsection.

\subsection{Gaussian process regression}~\label{sec:method}
A Gaussian Process (GP) $f(x)$ is a stochastic process which describes the properties of functions in function space using Gaussian distributions~\cite{Rasmussen:2006gp}. 
Given a set of $N$ data points $\{x_i,y_i\}, 1\leq i \leq N$ and a covariance matrix of the data, $\Sigma$, this formalism can be used as a nonparametric regression tool to estimate the underlying function via the mean $\bar{f}(x)$ and its uncertainty via the covariance $\mathrm{cov}(f)$.

In the following, we will identify $\mathbf{y}$ with the vector $\mathbf{\mathcal{C}_\ell}$ formed by the binned $\mathcal{C}_\ell$ of the {\it Planck} \texttt{plik\_lite} likelihood, $\mathbf{x}$ with the vector $\mathbf{L}$ of the weighted averages $\bar{\ell}_i$ of the multipoles included in bin~$i$ and $\Sigma$ with the corresponding covariance matrix.

In addition to the data vectors and the covariance matrix, we need to specify the mean of the Gaussian process $\textbf{m}$, which can in general be a function of $\ell$, and a kernel or covariance function $\text{cov}(f(\ell), f(\ell')) \equiv k(\ell,\ell')$.

The Gaussian process evaluated at multipoles $\mathbf{L}'$ follows the joint probability distribution given by~\cite{Rasmussen:2006gp, Arman2012gp}

\begin{equation}
  \label{eq:gp_matrix}
\begin{bmatrix}
\mathbfcal{C}_\mathbf{\ell}\\ 
\textbf{f}
\end{bmatrix}=\mathcal {N} \left ( \begin{bmatrix}
\textbf{m}(\textbf{L})\\ 
\textbf{m}(\textbf{L}')
\end{bmatrix},
\begin{bmatrix}
K_y(\textbf{L}, \textbf{L})& K(\textbf{L},\textbf{L}')\\ 
K(\textbf{L}',\textbf{L}) & K(\textbf{L}',\textbf{L}')
\end{bmatrix} \right ),
\end{equation}
where $\mathcal{N}$ is a multivariate normal distribution, 
$K_y(\textbf{L},\textbf{L}) \equiv K(\textbf{L},\textbf{L}) + \Sigma$ with the entries of $K(\textbf{L},\textbf{L})$ given by \mbox{$[K(\textbf{L},\textbf{L})]_{ij}=k(\ell_i, \ell_j)$}.

The mean of $\textbf{f}$ given the data can then be expressed as
\begin{equation}
  \label{eq:f_bar}
\bar{\textbf{f}}=\textbf{m}(\textbf{L}')+K(\textbf{L}',\textbf{L}) K_y^{-1}(\textbf{L},\textbf{L})(\textbf{y}-\textbf{m}(\textbf{L}))
\end{equation}
with covariance
\begin{equation}
  \label{eq:cov_f}
\text{cov}(\textbf{f})=K(\textbf{L}',\textbf{L}')-K(\textbf{L}',\textbf{L})K_y^{-1}(\textbf{L},\textbf{L}) K(\textbf{L},\textbf{L}').
\end{equation}

We take a Gaussian kernel function,
\begin{equation}
  \label{eq:kernel}
k(\ell, \ell')=\sigma^2_f \; \text{exp} \left( - \frac{(\ell-\ell')^2}{2 l^2_f} \right),
\end{equation}
characterised by two \textit{hyperparameters}: firstly, the correlation length $l_f$, which determines the stiffness of the Gaussian process and sets the typical scale on which the resulting $\bar{\mathbf{f}}$ can vary significantly and secondly, the prior width $\sigma_f$, which limits the allowed deviation of $f$ from the mean function.\footnote{Note that in Ref.~\cite{Rasmussen:2006gp} the $l_f$ and $\sigma_f$ are called length-scale and signal variance, respectively.} The choice of hyperparameters strongly influences the outcome of the regression; for example, a correlation length much shorter than the typical distance of two data points might lead to an overfitted function with too much structure, whereas a correlation length much larger than the total range covered by the data could result in an $\bar{f}$ that fits the data very poorly.

However, instead of simply arbitrarily fixing the hyperparameters, one can make use of the \textit{marginal likelihood} to let the data decide on their most suitable values. The marginal likelihood is defined as the conditional probability of the data given a set of hyperparameters, marginalised over all realisations of the GP,
\begin{equation}
\label{eq:like}
\mathcal{L}(\mathcal{C}_\ell|\textbf{L},l_f,\sigma_f) \equiv \int \mathrm{d}\mathbf{f} \, \mathcal{L}(\mathbfcal{C}_\mathbf{\ell}|\textbf{L},\mathbf{f},l_f,\sigma_f) \, p(\mathbf{f}|\textbf{L}),
\end{equation}
with a Gaussian prior $p$, and can be written as
\begin{equation}
  \label{eq:ln_like}
  \ln \mathcal {L} = -\frac{1}{2}(\textbf{y}-\textbf{m}(\textbf{L}))^T K_y^{-1}(\textbf{L},\textbf{L})(\textbf{y}-\textbf{m}(\textbf{L}))-\frac{1}{2} \ln \mid K_y(\textbf{L},\textbf{L}) \mid-\frac{N}{2}\ln2 \pi.
\end{equation}
Here, we perform the GP regression with the combination of $\sigma_f$ and $l_f$ that maximises $\mathcal{L}(\sigma_f,l_f)$\footnote{Alternatively, in a more strictly Bayesian spirit, one could also consider marginalising $f$ over the space of hyperparameters, weighted with $\mathcal{L}$, i.e., $f_\mathrm{marg} \propto \int \mathrm{d}\sigma_f \mathrm{d}l_f f(\sigma_f,l_f) \mathcal{L}(\sigma_f,l_f).$}.

\subsection{Using GP regression as a consistency test \label{sec:GPconsistency}}
In fact, the marginal likelihood $\mathcal{L}(\sigma_f,l_f)$ holds additional useful information that allows us to construct a consistency test: the location of the maximum can tell us whether the data prefer extra structure that is not present in the assumed mean function. 

Let us begin by discussing the expected form of $\mathcal{L}(\sigma_f,l_f)$ in the absence of extra structure: consider first the limiting case where one chooses the mean function to go exactly through all data points, i.e., $\mathbf{m(L)} = \mathbfcal{C}_\mathbf{\ell}$. In this case, any deviation from $\mathbf{m(L)}$ will decrease the likelihood, and thus for fixed $l_f$, the marginal likelihood will punish the addition of "extra covariance" and prefer a kernel function (Equation~(\ref{eq:kernel})) equal to zero.  At the same time, for fixed $\sigma_f$, it will prefer a stiffer GP over a more pliable one.  In other words, the global maximum of $\mathcal{L}$ will be at $\{\sigma_f\} \rightarrow \{0\}$.  For data which do scatter around the background but are still consistent with the background function, the global maximum of $\mathcal{L}(\sigma_f,l_f)$ may be a shallow bump at finite values of $\sigma_f$ and $l_f$ instead.

Conversely however, if $\mathcal{L}$ has a sufficiently pronounced peak, this may hint at a discrepancy between the data and the mean function, with the best-fit $\{\sigma_f,l_f\}$ indicating the typical amplitude and scale of the deviation, respectively.  Thus, if we set the mean function to be the best-fit of a particular model to the data, an inspection of $\mathcal{L}(\sigma_f,l_f)$ can reveal the presence of hidden patterns in the data and hence serve as a consistency test between model and data~\cite{Arman2013gp}.  

Since our goal is to probe the consistency of \textit{Planck} data with the $\Lambda$CDM model, it is natural to set the mean function to the respective base $\Lambda$CDM best-fit angular power spectra.  In particular, we use the primary $\mathcal{C}_\ell$ that maximise the \textit{Planck}~TTTEEE+\texttt{lowP} likelihood, downloaded from the \textit{Planck} Legacy Archive.\footnote{Since the 2015 \textit{Planck} data release, results from a new low-$\ell$ polarisation likelihood (\texttt{SimLow}) have been presented by the \textit{Planck} collaboration in Ref.~\cite{Aghanim:2016yuo}, but the neither the likelihood nor the best-fit parameters or spectra are in the public domain at the time of writing.  
Despite the significant shift in the inferred optical depth to reionisation $\tau$, there appears to be no degradation to the quality of the fit to the high-$\ell$ \textit{Planck} data~\cite{Aghanim:2016yuo}.  We thus would not expect any qualitative changes to our results if we were to use the \textit{Planck}~TTTEEE+\texttt{SimLow} best-fit spectrum instead.}  
Note that this is equivalent to performing a GP regression with mean zero on the residuals of the data with respect to the base $\Lambda$CDM best fit.

\begin{figure}
\begin{center}
 \includegraphics[height=.49\columnwidth,angle=270]{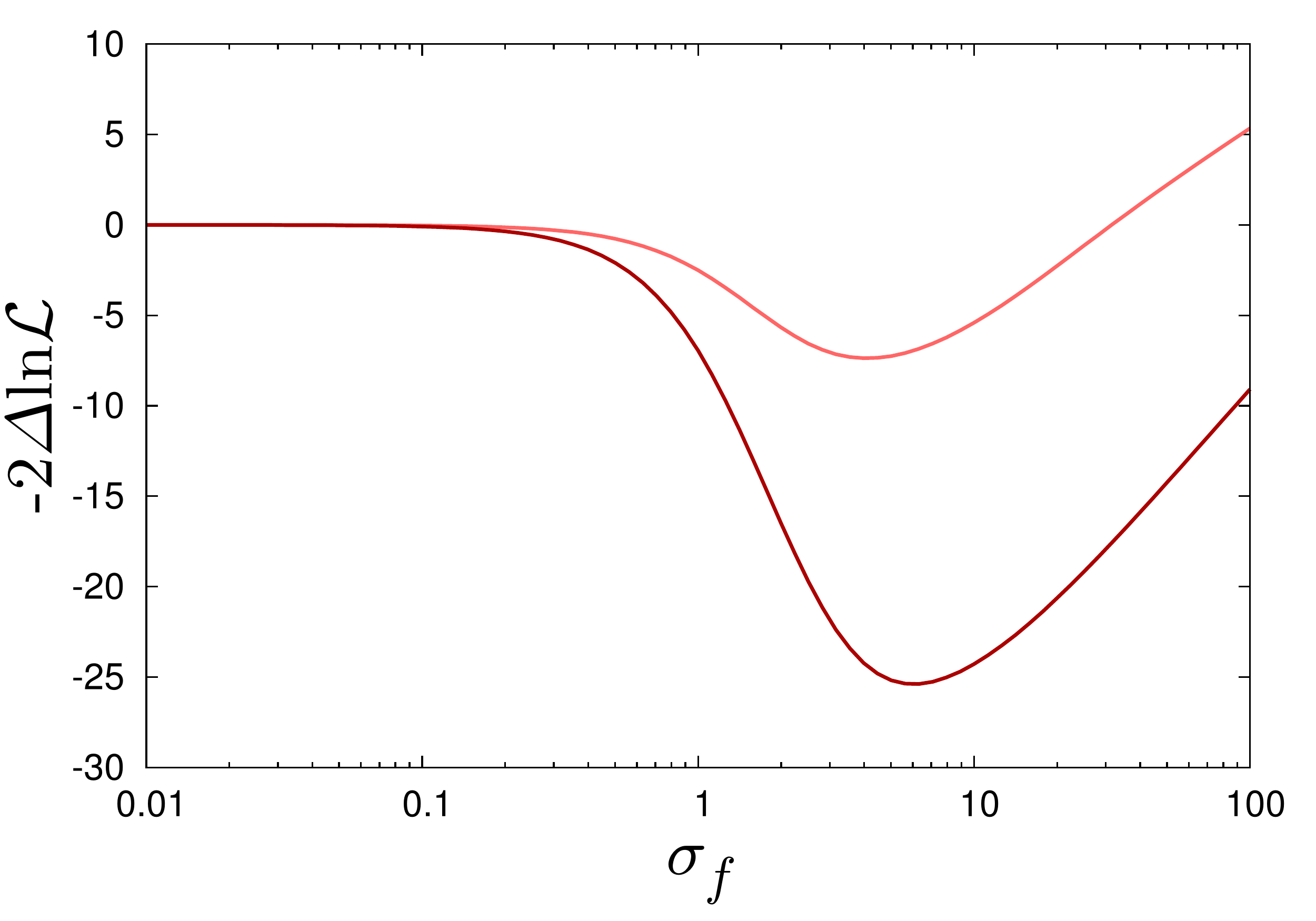}
 \includegraphics[height=.49\columnwidth,angle=270]{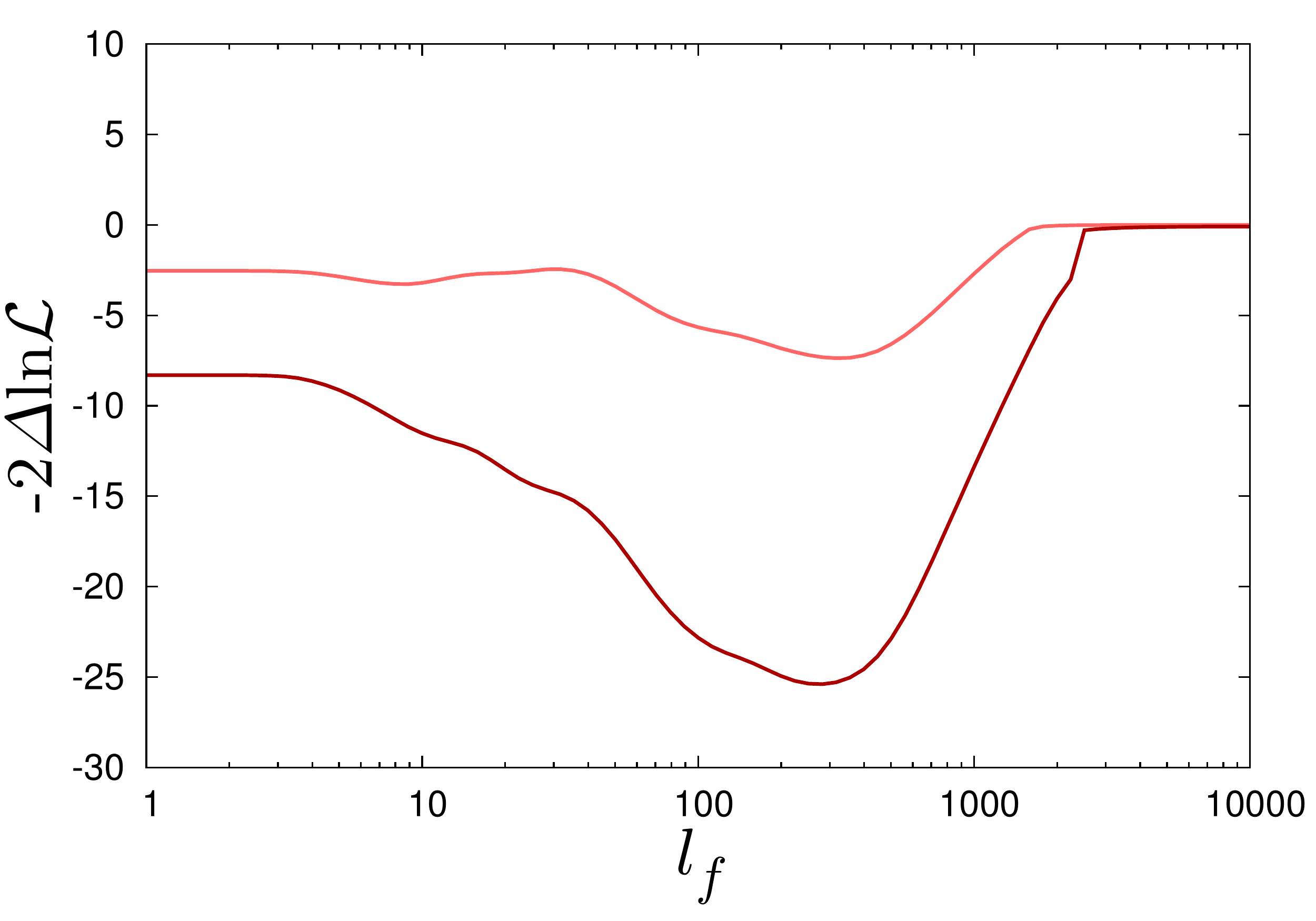}
 \caption{\label{fig:TTsim} One-dimensional profiles of the log-likelihood function in the $\sigma_f$- (\textit{left panel}) and $l_f$-direction (\textit{right panel}), if an artificial modulation signal of the form $\Delta \mathcal{C}_\ell = A \cos \left( 2\pi/\nu \; \ell \right)$ were added to the \textit{Planck} TT data.  In these figures, the modulation frequency is $\nu = 100$ and the amplitude $A=5 \, \mu \mathrm{K}^2$ (light red lines) or $A=8 \, \mu \mathrm{K}^2$ (dark red lines), respectively.  The log-likelihoods are normalised with respect to the $\Lambda$CDM best-fit.}
 \end{center}
\end{figure}

In Figure~\ref{fig:TTsim} we illustrate the ability of this method to detect a hidden signal.  In this particular example, we modulate the \textit{Planck} TT data with a cosine function (see the lower panel of the top plot in Figure~\ref{fig:residuals_GP}).  Even for a relatively modest modulation amplitude of $8 \, \mu \mathrm{K}^2$, the effective $\Delta \chi^2 \equiv - 2 \Delta \ln \mathcal{L} \equiv - 2 \left(\ln \mathcal{L}(\sigma_f,l_f) - \ln \mathcal{L}_{\Lambda\mathrm{CDM}}\right)$ is smaller than $-25$, providing a clear signal.  We can also observe the typical limiting behaviour of $\Delta \chi^2 \rightarrow 0$ as $\sigma_f \rightarrow 0$ and $l_f \rightarrow \infty$.

\subsection{Using the Bayesian evidence for model comparison}
When it comes to deciding whether or not a GP-improved model is statistically preferred over the fit of the mean function to the data, one can  take the idea behind Equation~(\ref{eq:like}) one step further, and marginalise over the hyperparameter directions as well, leading to the Bayesian evidence,
\begin{equation}
\label{eq:evidence}
\mathcal{E}_\mathrm{GP} = \int \mathrm{d}\sigma_f \mathrm{d}l_f \; \tilde{p}(\sigma_f,l_f) \mathcal{L}(\sigma_f,l_f),
\end{equation}
where one needs to specify a prior probability density $\tilde{p}$.  This is to be compared to the evidence for the background model $\mathcal{E}_\mathrm{bg}$, which, in this case, is equivalent to the likelihood of the data for the base $\Lambda$CDM best-fit).  Assuming equal model-priors, the Bayes factor
\begin{equation}
\label{eq:evidence2}
B \equiv \frac{\hat{\mathcal{E}}_\mathrm{GP}}{\hat{\mathcal{E}}_{bg}}
\end{equation}
quantifies the relative probabilities of the two models~\cite{Trotta:2008qt}, with $\ln B > 0$ indicating a preference for the GP-model.\footnote{Note that in our case we can exploit the fact that the models are {\it nested}, since the GP model reduces to the background model in the limit $\sigma_f \rightarrow 0$.  It is thus not necessary to perform the integration over $\Lambda$CDM parameter space $\mathbf{\Theta}$ to calculate the full evidences $\hat{\mathcal{E}}_\mathrm{GP} \equiv \int \mathrm{d}\mathbf{\Theta} p(\Theta) \mathcal{E}_\mathrm{GP}(\Theta)$ and $\hat{\mathcal{E}}_{bg} \equiv \int \mathrm{d}\mathbf{\Theta} p(\Theta) \mathcal{L}(\mathbf{\Theta})$; one can instead evaluate $B$ through the Savage-Dickey density ratio~\cite{Trotta:2008qt}, i.e., $B = \left. \frac{\mathcal{E}_\mathrm{GP}}{\mathcal{L}(\sigma_f,l_f)} \right|_{\sigma_f = 0}$.}  We do emphasize that the Bayes factor can have a considerable dependence on the choice of $\tilde{p}$, and should be taken with a grain of salt.  In the following, we will assume top-hat priors on the logarithms of the hyperparameters with the (data set dependent) prior ranges listed in Table~\ref{tab:priors}.

\begin{table}[tbp]
\centering
\begin{tabular}{c|ccc}
 & TT & TE & EE \\ \hline
$\log_{10} \sigma_f$ & $[ -2, 2]$ & $[ -2, 1.2]$ & $[ -2, 1.5]$  \\
$\log_{10} l_f$ & $[ 0,4 ]$ & $[ 0,4 ]$ & $[ 0,4 ]$ 
\end{tabular}
\caption{Prior ranges for the hyperparameters.
\label{tab:priors}}
\end{table}

In the fake-signal example of Section~\ref{sec:GPconsistency}, these priors lead to Bayes factors of $\ln B = 0.55$ for a modulation amplitude of $A=5 \, \mu \mathrm{K}^2$ (a very mild preference for the GP-model), and $\ln B = 8.3$ for the case of $A=8 \, \mu \mathrm{K}^2$ (decisive evidence for the GP-model).

\section{Results and Discussion \label{sec:results_discussion}}
\begin{figure}[ht]
\begin{center}
 \includegraphics[height=.49\columnwidth,angle=270]{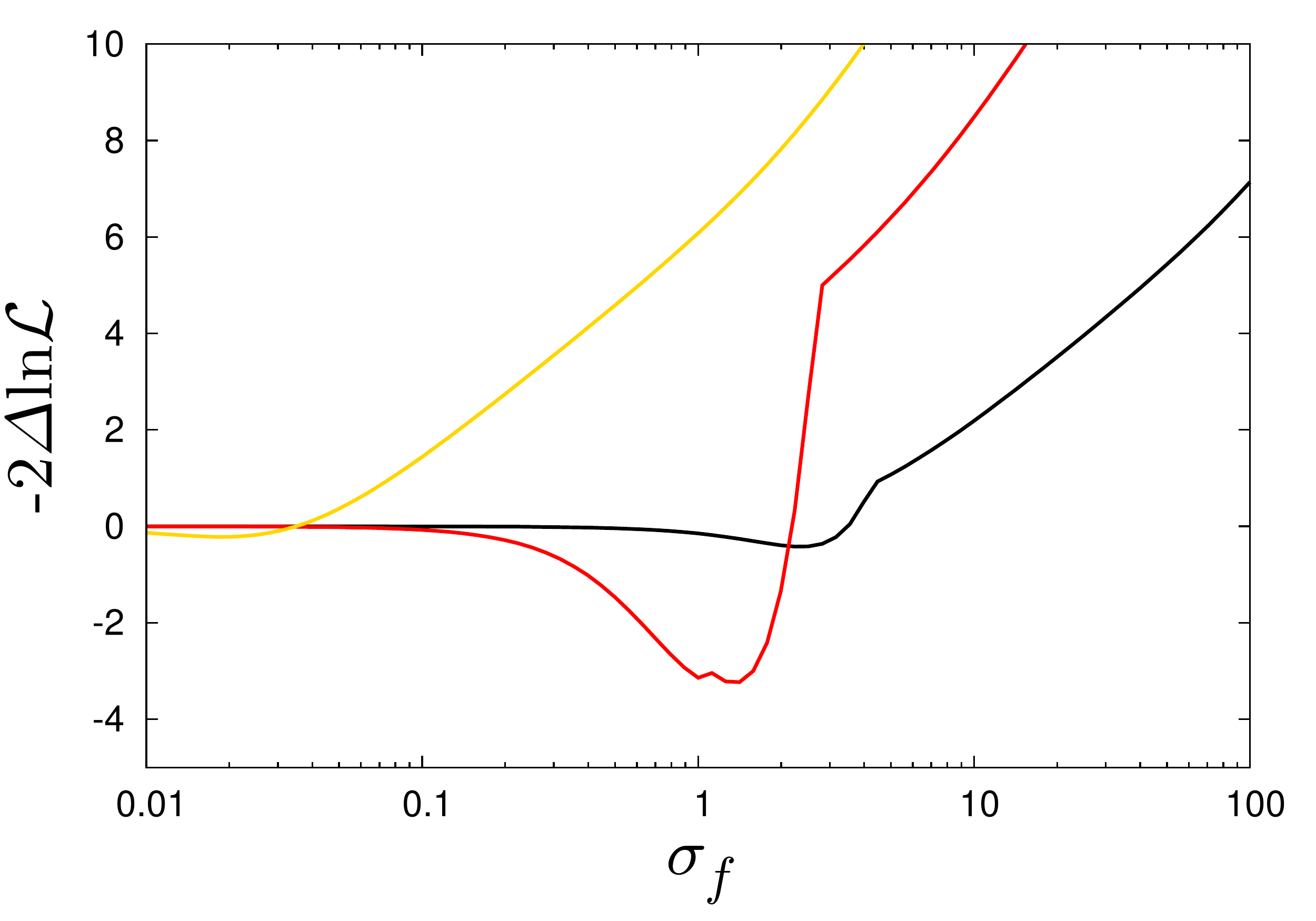}
 \includegraphics[height=.49\columnwidth,angle=270]{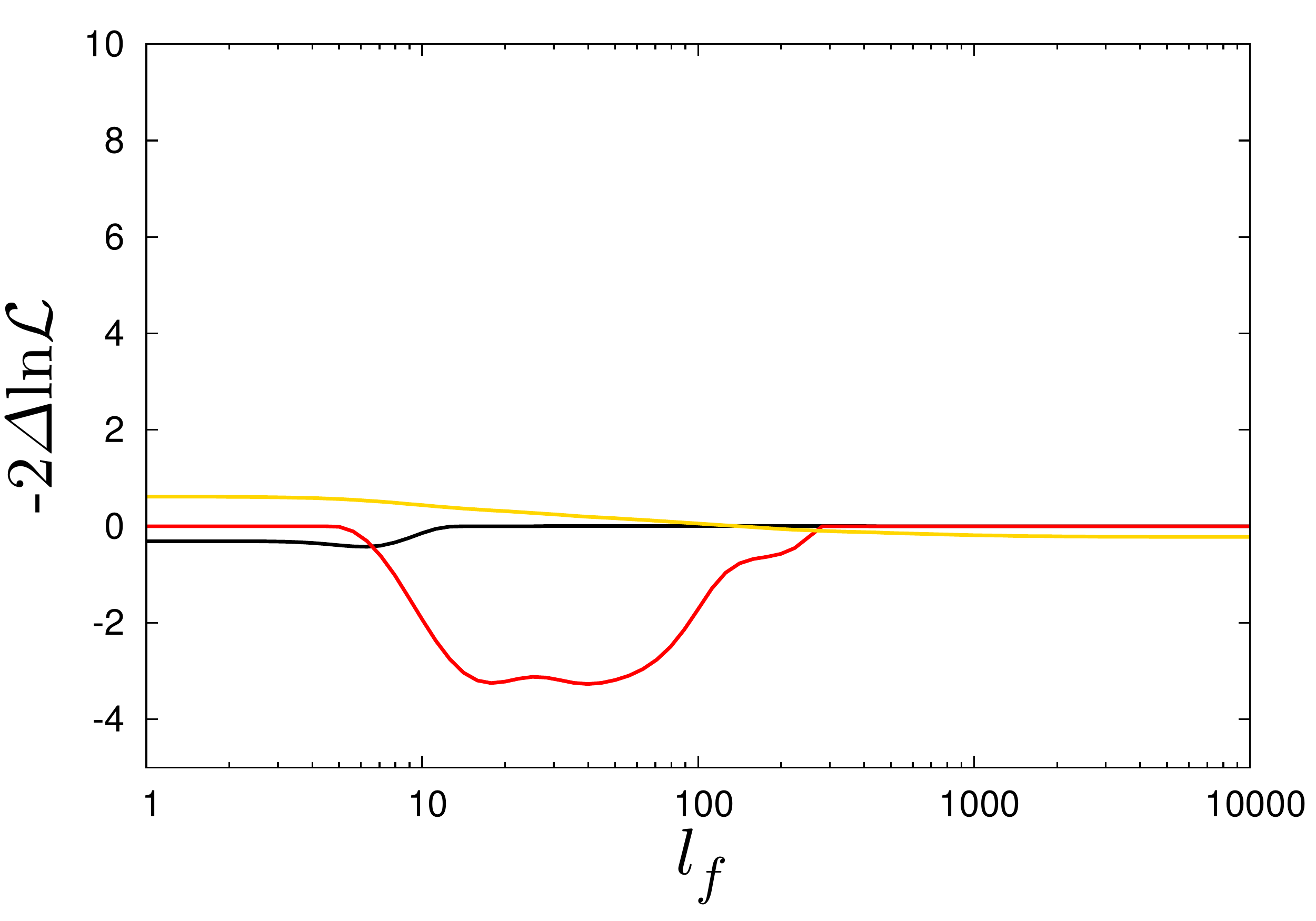}
 \caption{\label{fig:1dprofiles} One-dimensional profiles of the log-likelihood function in the $\sigma_f$- (\textit{left panel}) and $l_f$-direction (\textit{right panel}), for the \textit{Planck} TT (\textit{black}), TE (\textit{red}) and EE (\textit{gold}) data.  Just like in Figure~\ref{fig:TTsim}, the log-likelihoods are normalised with respect to the $\Lambda$CDM best-fit.}
 \end{center}
\end{figure}

In Figure~\ref{fig:1dprofiles}, we show the profiles of the logarithm of the marginal likelihood in the  $\sigma_f$- and $l_f$-directions for the three sets of angular power spectrum residuals.  For the best-fit hyperparameters, we plot the resulting mean and 1- and 2-$\sigma$ uncertainty bands of the Gaussian process, along with the residuals of the data, in Figure~\ref{fig:residuals_GP}.  As can be seen from Equation~(\ref{eq:cov_f}), if the covariance of the data $\Sigma$ dominates over the inherent covariance of the GP $K$, as is the case for the TT and EE data, the variance of the GP is given by $\sigma_f^2$.  In other words, for these particular choices of hyperparameters, the data are not able to meaningfully constrain the GP beyond its original prior width, which might be a consequence of using a constant $\sigma_f$ (as opposed to a function $\sigma_f(\ell)$) for data with a strongly $\ell$-dependent variance.
This is not the case for TE, where the size of the error bars is more uniform over the range of $\ell$ spanned by the data, and hence the variance of the GP is smaller than $\sigma_f^2$ here.

\begin{figure}[ht]
\begin{center}
 \includegraphics[height=.57\columnwidth,angle=270]{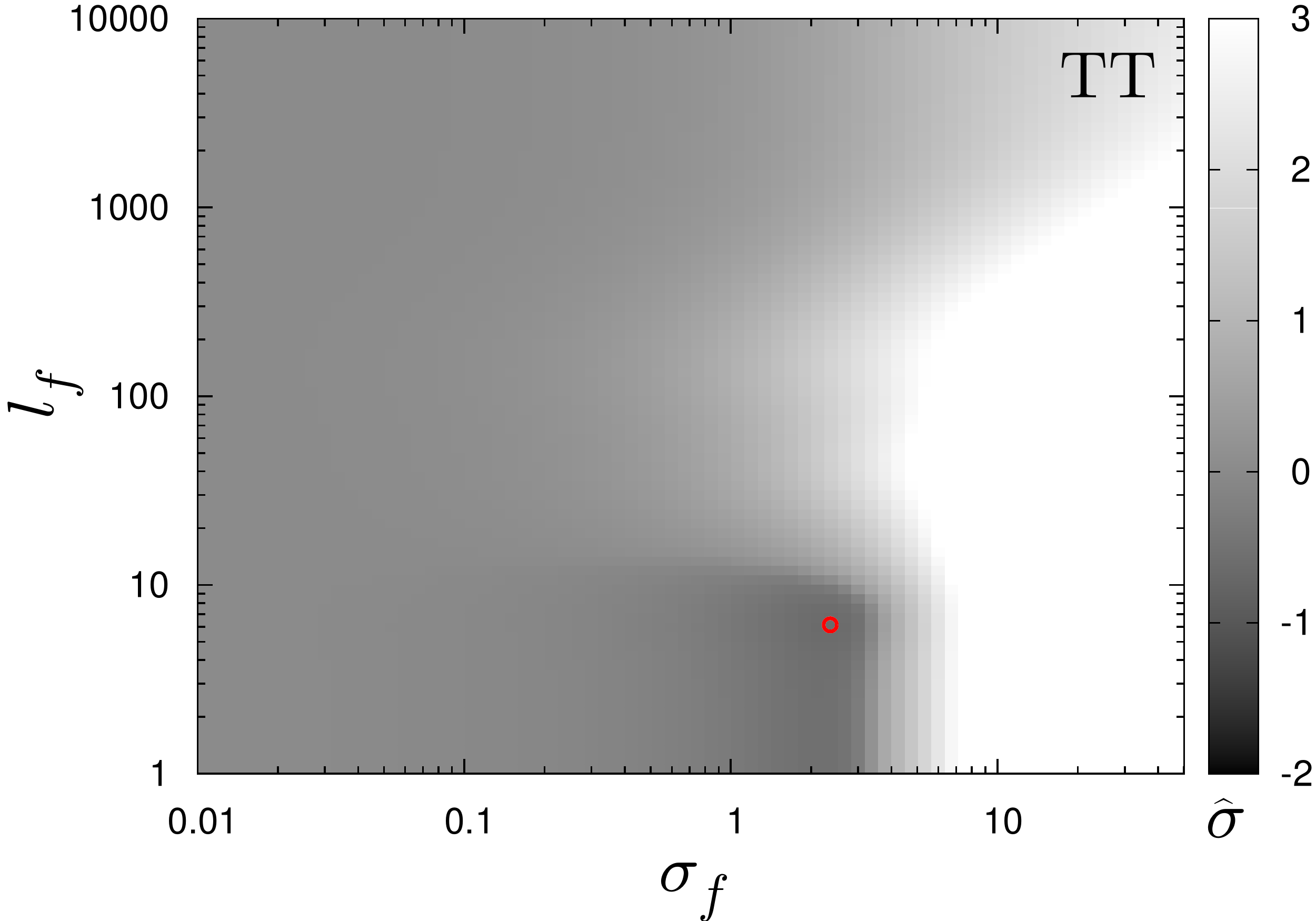}
 \includegraphics[height=.57\columnwidth,angle=270]{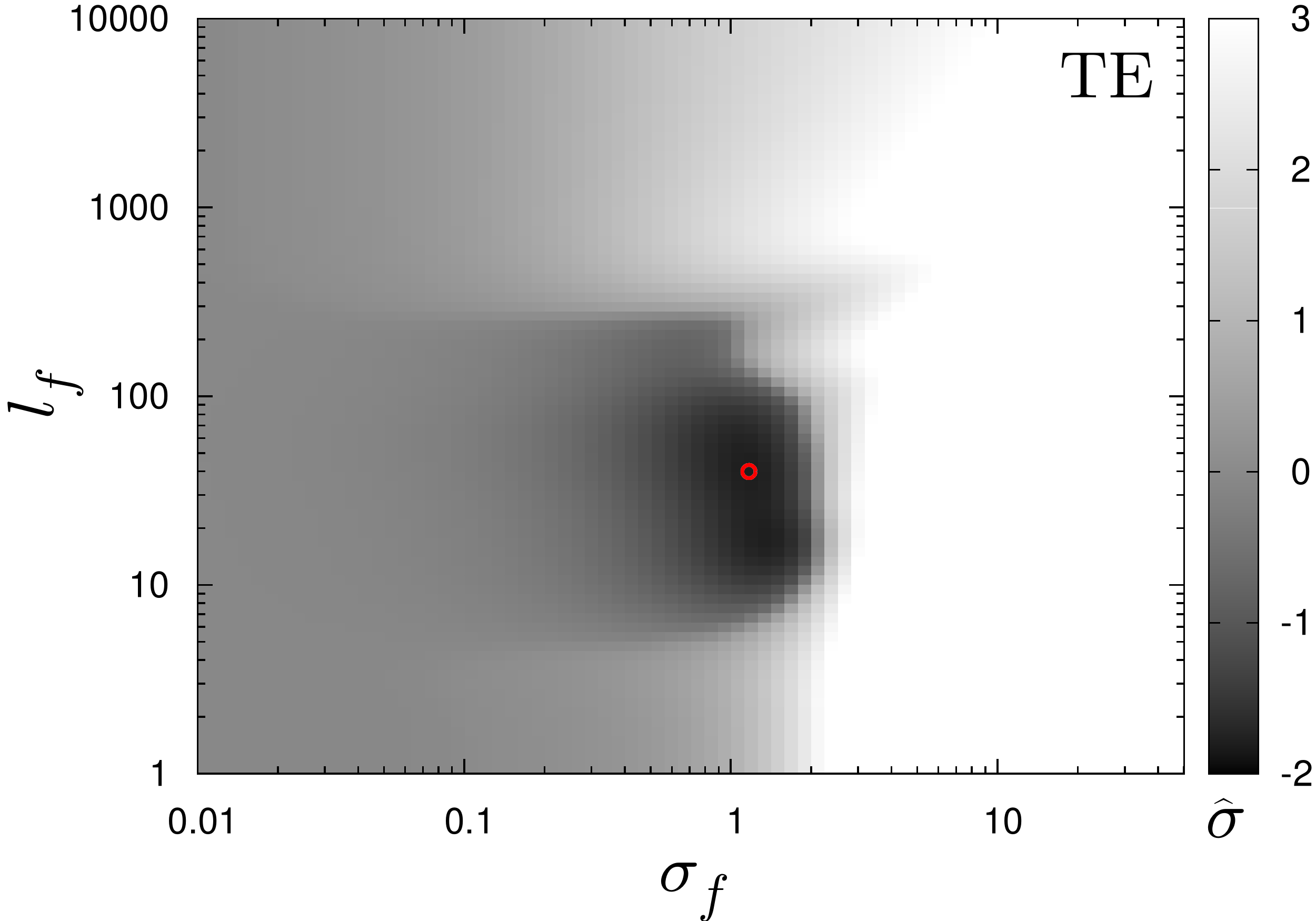}
 \includegraphics[height=.57\columnwidth,angle=270]{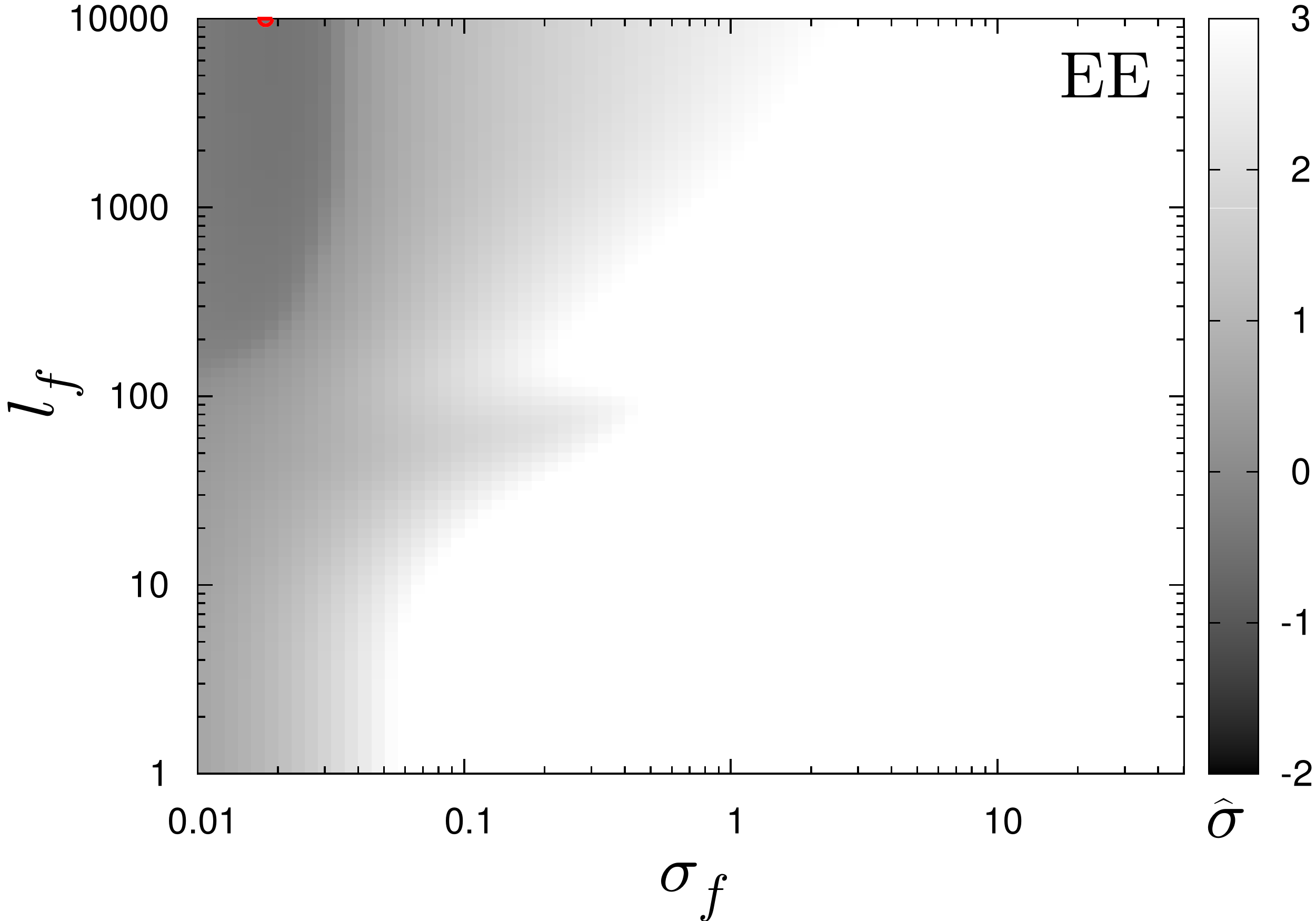}
 \caption{\label{fig:likelihood2d} This figure illustrates the dependence of the marginal likelihood on the hyperparameters $l_f$ and $\sigma_f$ for the \textit{Planck} TT (\textit{top}), TE (\textit{centre}) and EE (\textit{bottom}) data.  The colour scale shows the quantity $\hat{\sigma} \equiv \mathrm{sgn}(\Delta \chi^2) \cdot \sqrt{|\Delta \chi^2|}$ where $\Delta \chi^2 \equiv - 2 \Delta \ln \mathcal{L} \equiv - 2 \left(\ln \mathcal{L}(\sigma_f,l_f) - \ln \mathcal{L}_{\Lambda\mathrm{CDM}}\right)$. A negative $\hat{\sigma}$ corresponds to regions of parameter space which yield a better marginal likelihood than the base $\Lambda$CDM model best fit.  The optimal values of the hyperparameters are marked by red circles. }
\end{center}
\end{figure}

We note that the minima of the projection of $-2 \ln \mathcal{L}$ in the $l_f$-direction do not lie at $\sigma_l = 0$, but at $(\sigma_f,l_f) = (2.4,6.0)$ for TT, $(1.4,40)$ for TE and $(0.019,\infty)$ for EE (see also Figure~\ref{fig:likelihood2d} for the joint 2-dimensional likelihood). Taken at face value, this might suggest some preference for additional covariance in the residuals.
However, when dealing with real data which scatter around the true model, this should not come as a surprise.  We find that effective $\Delta \chi^2$ of the optimal Gaussian process regression with respect to the base $\Lambda$CDM best fit is $\Delta \chi^2_\mathrm{eff} = -0.4$ for the TT data, $\Delta \chi^2_\mathrm{eff} = -3.2$ for TE and $\Delta \chi^2_\mathrm{eff} = -0.2$ for EE, respectively.  Such mild improvements in the marginal likelihood cannot be interpreted as evidence for extra structure and are likely to be due to the stochastic spread of the data.

And indeed, for our choice of prior ranges, the Bayesian evidence does not show any preference for the GP-model either, with $\ln B^\mathrm{TT} = -0.5$ for the TT data, $\ln B^\mathrm{TE} = -0.3$ for the TE data and $\ln B^\mathrm{EE} = -1.6$ for the EE data.

In the case of the TT data, this conclusion is further supported by three observations that are related to the smallness of $\Delta \chi_\mathrm{eff}^2$: firstly, the optimal values of $l_f$ are $\mathcal{O}(10)$, comparable to the average bin size. Secondly, the amplitude of the modulation is much smaller than the mean standard deviation of the residuals (cf.~Figure~\ref{fig:residuals_GP}).  Thirdly, the maximum local deviation of the GP from the mean does not exceed one standard deviation, with the global significance being even smaller due to the look-elsewhere effect. 
The latter two points are also true for the EE case, but here, instead of adding structure, the GP prefers a nearly constant offset, albeit with an amplitude that is completely negligible when compared to both the amplitude of the EE power spectrum and the standard deviation of the residuals. We should note that a similar constant amplitude offset was reported recently by~\cite{Shafieloo:2016zga}. While the mean amplitude of the offset in this analysis seems to be smaller than the one found in~\cite{Shafieloo:2016zga}, results of the two analyses are consistent considering the uncertainties. This effect could potentially be caused by a permille-sized mismatch in the calibration of the polarisation spectra with respect to the temperature spectra, or might also be related to the absence of the low-$\ell$ \textit{Planck} data in our approach (note that the base $\Lambda$CDM best-fit spectra we assume are the best-fits with respect to the full \textit{Planck} likelihood).

Things are slightly more interesting for the TE data: the preferred correlation length is clearly larger than the average bin size and the optimal prior width reaches about 22\% of the average standard deviation of the residuals.  Nonetheless, neither the local deviation of the GP from the base $\Lambda$CDM best-fit nor the absolute $\Delta \chi^2_\mathrm{eff}$ indicate a serious discrepancy.  A likely reason for this could be the fact that the base $\Lambda$CDM model's best-fit to TE only does not completely coincide with the full TTTEEE data fit, which is still dominated by the TT data (explaining why the TT data GP does not display the same behaviour).  But at the same time, the TE data by themselves are starting to become competitive with the TT data for parameter estimation purposes~\cite{Ade:2015xua}, so unlike the EE data, they are sensitive enough for this small difference in best-fits to actually be picked up by the GP regression test.

\section{Conclusion\label{sec:conclusion}}
We have tested the consistency of \textit{Planck} TT, TE and EE angular power spectrum data with the base $\Lambda$CDM model using a non-parametric test based on Gaussian process regression.  Being non-parametric and applied directly to the observable (rather than a fundamental quantity, e.g., the primordial power spectrum), this method is sensitive to general inconsistencies between data and model, no matter whether they are caused by assuming a wrong physical model or unknown systematics in the data.

It should be pointed out that since we used binned data, the scope of our analysis is limited to structures with a characteristic variation length of the order of the respective bin size.  It would be interesting to repeat the analysis on unbinned data in order to increase sensitivity to finer structures with $\Delta \ell \sim 1$, but unfortunately, \textit{Planck}'s frequency-combined, foreground-marginalised power spectra are not publically available in unbinned form at present.

Nevertheless, our results do not show any serious inconsistencies; besides a statistically non-significant deviation in the TE spectrum, we find excellent agreement between \textit{Planck} data and the best-fit base $\Lambda$CDM spectra: especially the TT residuals display a perhaps remarkable lack of hidden patterns.  We therefore conclude that the $\Lambda$CDM model passes yet another challenge thrown at it, and confirm its consistency with \textit{Planck} data.

\section{Acknowledgments}~\label{sec:acknowledge}
J.H. gratefully acknowledges support of an Australian Research Council Future Fellowship  (FT140100481) and would like to thank Aarhus University and the Korea Astronomy and Space Science Institute for hospitality during the preparation of this work. A.S. would like to acknowledge the support of the National Research Foundation of Korea (NRF-2016R1C1B2016478) and Sydney Institute for Astronomy, University of Sydney for hospitality during the preparation of this work.

\bibliographystyle{JHEP}
\bibliography{references.bib}

\providecommand{\href}[2]{#2}\begingroup\raggedright\begin{thebibliography}{10}

\bibitem{Adam:2015rua}
{\scshape Planck} collaboration, R.~Adam et~al., \emph{{Planck 2015 results. I.
  Overview of products and scientific results}},
  \href{http://dx.doi.org/10.1051/0004-6361/201527101}{\emph{Astron.
  Astrophys.} {\bfseries 594} (2016) A1},
  [\href{https://arxiv.org/abs/1502.01582}{{\ttfamily 1502.01582}}].

\bibitem{Aghanim:2015xee}
{\scshape Planck} collaboration, N.~Aghanim et~al., \emph{{Planck 2015 results.
  XI. CMB power spectra, likelihoods, and robustness of parameters}},
  \href{http://dx.doi.org/10.1051/0004-6361/201526926}{\emph{Astron.
  Astrophys.} {\bfseries 594} (2016) A11},
  [\href{https://arxiv.org/abs/1507.02704}{{\ttfamily 1507.02704}}].

\bibitem{Ade:2015xua}
{\scshape Planck} collaboration, P.~A.~R. Ade et~al., \emph{{Planck 2015
  results. XIII. Cosmological parameters}},
  \href{http://dx.doi.org/10.1051/0004-6361/201525830}{\emph{Astron.
  Astrophys.} {\bfseries 594} (2016) A13},
  [\href{https://arxiv.org/abs/1502.01589}{{\ttfamily 1502.01589}}].

\bibitem{Ade:2015lrj}
{\scshape Planck} collaboration, P.~A.~R. Ade et~al., \emph{{Planck 2015
  results. XX. Constraints on inflation}},
  \href{http://dx.doi.org/10.1051/0004-6361/201525898}{\emph{Astron.
  Astrophys.} {\bfseries 594} (2016) A20},
  [\href{https://arxiv.org/abs/1502.02114}{{\ttfamily 1502.02114}}].

\bibitem{Ade:2015rim}
{\scshape Planck} collaboration, P.~A.~R. Ade et~al., \emph{{Planck 2015
  results. XIV. Dark energy and modified gravity}},
  \href{http://dx.doi.org/10.1051/0004-6361/201525814}{\emph{Astron.
  Astrophys.} {\bfseries 594} (2016) A14},
  [\href{https://arxiv.org/abs/1502.01590}{{\ttfamily 1502.01590}}].

\bibitem{Rasmussen:2006gp}
C.~Rasmussen and C.~Williams, \emph{Gaussian Processes for Machine Learning}.
\newblock Adaptative computation and machine learning series. University Press
  Group Limited, 2006.

\bibitem{Heitmann2010a}
T.~{Holsclaw}, U.~{Alam}, B.~{Sans{\'o}}, H.~{Lee}, K.~{Heitmann}, S.~{Habib}
  et~al., \emph{{Nonparametric reconstruction of the dark energy equation of
  state}}, \href{http://dx.doi.org/10.1103/PhysRevD.82.103502}{\emph{Physical
  Review D} {\bfseries 82} (Nov., 2010) 103502},
  [\href{https://arxiv.org/abs/1009.5443}{{\ttfamily 1009.5443}}].

\bibitem{Heitmann2010b}
T.~{Holsclaw}, U.~{Alam}, B.~{Sans{\'o}}, H.~{Lee}, K.~{Heitmann}, S.~{Habib}
  et~al., \emph{{Nonparametric Dark Energy Reconstruction from Supernova
  Data}},
  \href{http://dx.doi.org/10.1103/PhysRevLett.105.241302}{\emph{Physical Review
  Letters} {\bfseries 105} (Dec., 2010) 241302},
  [\href{https://arxiv.org/abs/1011.3079}{{\ttfamily 1011.3079}}].

\bibitem{Heitmann2011}
T.~{Holsclaw}, U.~{Alam}, B.~{Sans{\'o}}, H.~{Lee}, K.~{Heitmann}, S.~{Habib}
  et~al., \emph{{Nonparametric reconstruction of the dark energy equation of
  state from diverse data sets}},
  \href{http://dx.doi.org/10.1103/PhysRevD.84.083501}{\emph{Physical Review D}
  {\bfseries 84} (Oct., 2011) 083501},
  [\href{https://arxiv.org/abs/1104.2041}{{\ttfamily 1104.2041}}].

\bibitem{Arman2012gp}
A.~{Shafieloo}, A.~G. {Kim} and E.~V. {Linder}, \emph{{Gaussian process
  cosmography}},
  \href{http://dx.doi.org/10.1103/PhysRevD.85.123530}{\emph{Physical Review D}
  {\bfseries 85} (June, 2012) 123530},
  [\href{https://arxiv.org/abs/1204.2272}{{\ttfamily 1204.2272}}].

\bibitem{Seikel2012gp}
M.~{Seikel}, C.~{Clarkson} and M.~{Smith}, \emph{{Reconstruction of dark energy
  and expansion dynamics using Gaussian processes}},
  \href{http://dx.doi.org/10.1088/1475-7516/2012/06/036}{\emph{Journal of
  Cosmology and Astroparticle Physics} {\bfseries 6} (June, 2012) 036},
  [\href{https://arxiv.org/abs/1204.2832}{{\ttfamily 1204.2832}}].

\bibitem{Aghamousa2015b}
A.~{Aghamousa}, A.~{Shafieloo}, M.~{Arjunwadkar} and T.~{Souradeep},
  \emph{{Unveiling acoustic physics of the CMB using nonparametric estimation
  of the temperature angular power spectrum for Planck}},
  \href{http://dx.doi.org/10.1088/1475-7516/2015/02/007}{\emph{Journal of
  Cosmology and Astroparticle Physics} {\bfseries 2} (Feb., 2015) 007},
  [\href{https://arxiv.org/abs/1412.3552}{{\ttfamily 1412.3552}}].

\bibitem{Aghamousa2015c}
A.~{Aghamousa} and A.~{Shafieloo}, \emph{{Nonparametric test of consistency
  between cosmological models and multiband CMB measurements}},
  \href{http://dx.doi.org/10.1088/1475-7516/2015/06/003}{\emph{Journal of
  Cosmology and Astroparticle Physics} {\bfseries 6} (June, 2015) 003},
  [\href{https://arxiv.org/abs/1502.00851}{{\ttfamily 1502.00851}}].

\bibitem{Hazra:2014hma}
D.~K. Hazra and A.~Shafieloo, \emph{{Confronting the concordance model of
  cosmology with Planck data}},
  \href{http://dx.doi.org/10.1088/1475-7516/2014/01/043}{\emph{JCAP} {\bfseries
  1401} (2014) 043}, [\href{https://arxiv.org/abs/1401.0595}{{\ttfamily
  1401.0595}}].

\bibitem{Arman2013gp}
A.~{Shafieloo}, A.~G. {Kim} and E.~V. {Linder}, \emph{{Model independent tests
  of cosmic growth versus expansion}},
  \href{http://dx.doi.org/10.1103/PhysRevD.87.023520}{\emph{Physical Review D}
  {\bfseries 87} (Jan., 2013) 023520},
  [\href{https://arxiv.org/abs/1211.6128}{{\ttfamily 1211.6128}}].

\bibitem{Aghanim:2016yuo}
{\scshape Planck} collaboration, N.~Aghanim et~al., \emph{{Planck intermediate
  results. XLVI. Reduction of large-scale systematic effects in HFI
  polarization maps and estimation of the reionization optical depth}},
  \href{http://dx.doi.org/10.1051/0004-6361/201628890}{\emph{Astron.
  Astrophys.} {\bfseries 596} (2016) A107},
  [\href{https://arxiv.org/abs/1605.02985}{{\ttfamily 1605.02985}}].

\bibitem{Trotta:2008qt}
R.~Trotta, \emph{{Bayes in the sky: Bayesian inference and model selection in
  cosmology}},
  \href{http://dx.doi.org/10.1080/00107510802066753}{\emph{Contemp. Phys.}
  {\bfseries 49} (2008) 71--104},
  [\href{https://arxiv.org/abs/0803.4089}{{\ttfamily 0803.4089}}].

\bibitem{Shafieloo:2016zga}
A.~Shafieloo and D.~K. Hazra, \emph{{Consistency of the Planck CMB data and
  $\Lambda$CDM cosmology}},
  \href{http://dx.doi.org/10.1088/1475-7516/2017/04/012}{\emph{JCAP} {\bfseries
  1704} (2017) 012}, [\href{https://arxiv.org/abs/1610.07402}{{\ttfamily
  1610.07402}}].

\end{thebibliography}\endgroup

\end{document}